\documentclass[aps,prl,reprint,superscriptaddress,showpacs]{revtex4-1}

\usepackage{graphicx}

\begin{document}


\title{Domain-wall induced large magnetoresistance effects at zero applied field in ballistic nanocontacts}


\author{Arndt von Bieren}
\affiliation{Laboratory for Nanomagnetism and Spin Dynamics, \'Ecole Polytechnique F\'ed\'erale de Lausanne, 1015 Lausanne, Switzerland and SwissFEL, Paul Scherrer Institut, 5232 Villigen PSI, Switzerland}
\affiliation{Fachbereich Physik, Universit\"at Konstanz, Universit\"atsstrasse 10, 78457 Konstanz, Germany}

\author{Ajit K. Patra}
\author{Stephen Krzyk}
\affiliation{Fachbereich Physik, Universit\"at Konstanz, Universit\"atsstrasse 10, 78457 Konstanz, Germany}

\author{Jan Rhensius}
\affiliation{Laboratory for Nanomagnetism and Spin Dynamics, \'Ecole Polytechnique F\'ed\'erale de Lausanne, 1015 Lausanne, Switzerland and SwissFEL, Paul Scherrer Institut, 5232 Villigen PSI, Switzerland}
\affiliation{Fachbereich Physik, Universit\"at Konstanz, Universit\"atsstrasse 10, 78457 Konstanz, Germany}
\affiliation{Laboratory for Micro- and Nanotechnology, Paul Scherrer Institut, 5232 Villigen PSI, Switzerland}

\author{Robert M. Reeve}
\affiliation{Institut f\"ur Physik, Johannes Gutenberg-Universit\"at Mainz, 55099 Mainz, Germany}

\author{Laura J. Heyderman}
\affiliation{Laboratory for Micro- and Nanotechnology, Paul Scherrer Institut, 5232 Villigen PSI, Switzerland}

\author{Regina Hoffmann-Vogel}
\affiliation{Physikalisches Institut and DFG-Center for Functional Nanostructures, Karlsruhe Institute of Technology Campus South, 76128 Karlsruhe, Germany}

\author{Mathias Kl\"aui}
\email{Klaeui@uni-mainz.de}
\affiliation{Laboratory for Nanomagnetism and Spin Dynamics, \'Ecole Polytechnique F\'ed\'erale de Lausanne, 1015 Lausanne, Switzerland and SwissFEL, Paul Scherrer Institut, 5232 Villigen PSI, Switzerland}
\affiliation{Fachbereich Physik, Universit\"at Konstanz, Universit\"atsstrasse 10, 78457 Konstanz, Germany}
\affiliation{Institut f\"ur Physik, Johannes Gutenberg-Universit\"at Mainz, 55099 Mainz, Germany}



\date{\today}


\begin{abstract}
We determine magnetoresistance effects in stable and clean permalloy nanocontacts of variable cross-section, fabricated by UHV deposition and in-situ electromigration. To ascertain the magnetoresistance (MR) effects originating from a magnetic domain wall, we measure the resistance values with and without such a wall at \textit{zero} applied field. 
In the ballistic transport regime, the MR ratio reaches up to 50\% and exhibits a previously unobserved sign change. Our results can be reproduced by recent atomistic 
calculations 
for different atomic configurations of the nanocontact, 
highlighting the importance of the detailed atomic arrangement for the MR effect.
\end{abstract}


\maketitle


The magneto-transport properties of a device change drastically when the dimensions become comparable to characteristic length scales, such as the mean free path, the Fermi wavelength or the exchange length \cite{agrait_quantum_2003}, and nanocontacts offer the possibility to study such length scales. Furthermore, magnetic nanocontacts can accommodate geometrically confined magnetic domain walls (DWs) as the wall width scales with the constriction width \cite{backes_transverse_2007}. In the smallest possible (i.e. atomic) contacts, this eventually leads to atomically abrupt spin structure changes.

In such narrow DWs the spins of the charge carriers can no longer adiabatically follow the spin structure direction. 
Consequently, significant magnetotransport effects have been predicted \cite{jacob_magnetic_2005,levy_resistivity_1997,tatara_resistivity_1997} and observed \cite{doudin_ballistic_2008,bolotin_ballistic_2006,sokolov_quantized_2007,keane_magnetoresistance_2006}, opening also 
the prospect of novel device applications 
\cite{terabe_quantized_2005}. However, reliable magnetoresistance (MR) measurements on nanocontacts necessitate 
particular requirements in terms of stability, cleanliness and control of the spin structure. These requirements have previously been unattainable \cite{garcia_magnetoresistance_1999,chopra_ballistic_2002} leading to the observation of artifacts such as magnetostriction \cite{doudin_ballistic_2008,egelhoff_artifacts_2004,muller_switching_2011} and contamination \cite{yoshida_importance_2009,untiedt_absence_2004}. 

\begin{figure}[b]
 \includegraphics{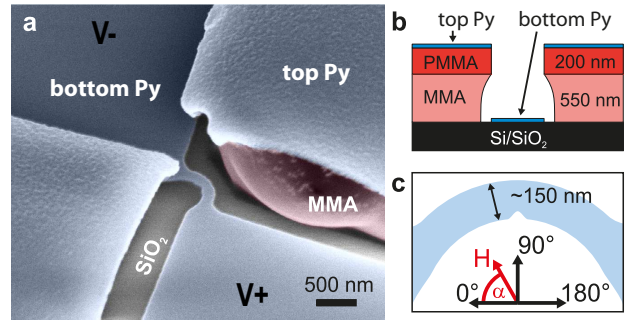}%
 \caption{\label{fig1} (Color online) Scanning electron microscope image (viewing angle $45^\circ$) showing the constricted ring section before electromigration (blue: Py, red: resist, gray: SiO$_2$). The deformation of the MMA/PMMA resist is caused by the electron beam of the SEM.  
(b) Schematic illustration of the cross-sectional view of the sample and (c) the top view of the ring section indicating the orientation of the in-plane magnetic field angle $\alpha$.}
 \end{figure}

In this Letter we report the first observation of large domain wall magnetoresistance (DWMR) effects up to 50\% at \textit{zero} field in exceptionally clean and stable permalloy (Py = Ni$_{80}$Fe$_{20}$) nanocontacts of tailored geometry \cite{klaeui_magnetization_2007}. In addition, the MR exhibits a previously unobserved sign change in the ballistic transport regime. This result can be reproduced by recent theoretical calculations highlighting the importance of the detailed atomic arrangement for the sign and magnitude of the MR \cite{achilles_tailoring_2011}.
The MR is measured 
at \textit{zero} applied field in nanocontacts that are rigidly attached to a substrate \cite{patra_magnetoresistance_2010}. Using a special device geometry, we control the spin structure and in particular the presence of a DW. Furthermore, the nanocontacts are fabricated and characterized at low temperatures in the same ultra-high vacuum (UHV) chamber without breaking the vacuum \cite{krzyk_magnetotransport_2010}. \emph{In-situ} electromigration allows us to reduce the cross-section of the nanocontacts in a controlled fashion during the study \cite{hoffmann_conductance_2008}. Using this unique approach, we study the MR in magnetic nanocontacts from the diffusive to the ballistic transport regime while minimizing artifacts due to magnetostriction and impurities.

After ex-situ fabrication of separated $5$ nm Ti / $50$ nm Au contact pads on a Si/SiO$_2$ substrate, a double layer resist mask 
was defined by electron beam lithography (see Fig. \ref{fig1}(b)). 
Then the sample was mounted on a chip carrier and the Au pads were 
electrically connected by wire bonding. 
In order to keep the contacts as clean and stable 
as possible all subsequent steps (i.e. deposition, electromigration and MR measurements) were carried out in-situ in the same ultra-high vacuum (UHV) chamber \cite{krzyk_magnetotransport_2010} 
with a base pressure of $3 \cdot 10^{-10}$ mbar. 
A 24 nm thick Py film was deposited onto the sample using thermal evaporation. 
The resulting patterned film on the SiO$_2$ surface (see Fig. \ref{fig1}(a)) 
connects the pairs of Au pads thus allowing for electrical characterization of the structure. Due to the large undercut along the edges of the resist, the contacted film is electrically isolated from the Py deposited on top of the resist, as can be seen in Fig. 1(a) and 1(b). 
To study the MR effects associated with DWs, we have chosen a magnetic half ring structure with a constriction at its center (shown in Fig. 1(a,c)). In this geometry, DWs can be positioned precisely and reproducibly using a rotatable in-plane magnetic field H (Fig. \ref{fig1}(c)) \cite{klaeui_domain_2003}. 

To obtain nanocontacts of different cross-section, we carry out successive automated electromigration of the half-ring wire. Electromigration is widely used to form constrictions with contacts down to the ballistic transport regime with quantized conduction \cite{bolotin_ballistic_2006, hoffmann_conductance_2008}. The constriction defines the position of the highest current density in the structure and it hence determines where electromigration sets in. 
A typical measurement cycle consists 
of electromigration, where the constriction is thinned, followed by the in-situ characterization of the MR \cite{patra_magnetoresistance_2010}. As the constriction is thinned, it quickly starts to dominate the overall resistance. It is hence primarily the MR response of this area which is probed by the MR measurements. 
Both electromigration and MR measurements are performed in UHV at a temperature of $80$ K. This drastically reduces thermal noise and allows us to obtain mechanically stable contacts, which is not possible at room temperature. 
The electromigration process is repeated until the contact is completely open, i.e., a gap has formed at the position of the constriction resulting in an open circuit. 
Due to the increased noise and reduced stability of the contacts in the tunneling regime above R $>50\ \mbox{k}\Omega$, we concentrate on stable resistances below that in the ballistic conduction regime. In this fashion we 
efficiently determine the evolution of the MR as a function of contact resistance, minimizing artifacts due to magnetostriction and impurities.

Three different types of measurement were employed to study the MR effects associated with DWs. The resistance of the contacts was measured (i) as a function of applied field angle $\alpha$ (see Fig. \ref{fig1}(c)) for a given applied field amplitude to determine the AMR, (ii) as a function of the field amplitude for a given field angle $\alpha$ (R(H)-loop) and (iii) as a function of field angle $\alpha$ at remanence after applying a magnetic field along $\alpha$ and reducing the field to zero (for details on (iii), see the quasi-static measurement scheme in Refs. \onlinecite{patra_magnetoresistance_2010,klaeui_domain_2003}). In the following, the MR ratio is defined as $(R_{AP} - R_{P}) / R_{P}$, where $R_{P}$ and $R_{AP}$ denote the resistance of the nanocontact with both arms of the half-ring oriented in a parallel (P state, no DW) and anti-parallel (AP state, DW at the constriction) configuration, respectively. 

 \begin{figure}[t]
 \includegraphics{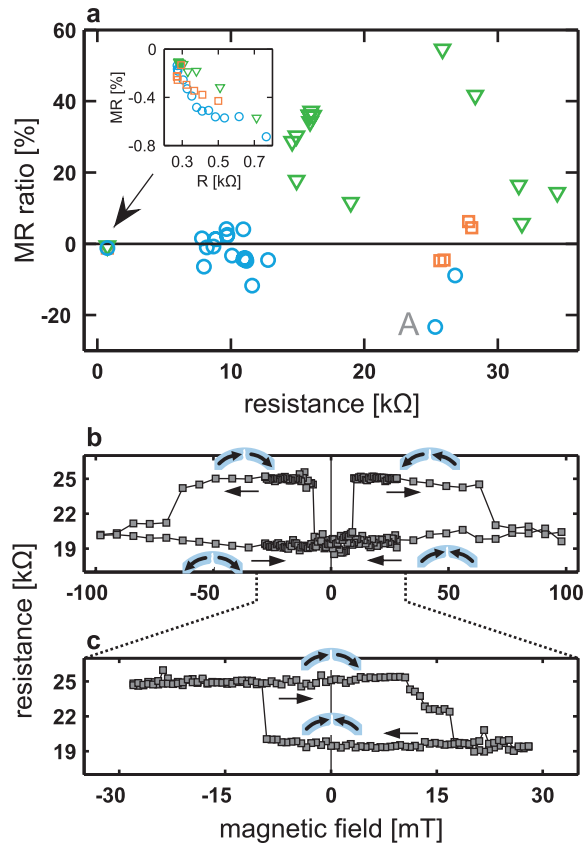}%
 \caption{\label{fig2} (Color online) (a) Magnetoresistance ratio $\mbox{MR} = (R_{AP}  R_P) / R_P$ vs. contact resistance in the parallel state ($R_P$) for three nanocontacts (blue, green and red). The data points are acquired from R(H) loops with $\mu_0 H_{max} = 100\ \mbox{mT}$ and $\alpha = 75^\circ$ and $90^\circ$. The inset shows the evolution of AMR for low contact resistances at the beginning of the electromigration process. Resistance vs. magnetic field (b) major and (c) minor loop at field angle $\alpha = 75^\circ$ for the contact state labeled `A' in Fig. 2(a). The sketches of the half-ring structure illustrate the magnetization configuration of the contact leads for different positions in the loop. The black arrows along the curve indicate the sweep direction.
}
\end{figure}

In agreement with our previous findings \cite{patra_magnetoresistance_2010,klaeui_domain_2003}, we observe different resistance levels with and without a DW pinned at the constriction in both the diffusive and the ballistic regime. The switching between these states 
is illustrated in Fig. \ref{fig2}(b): We first measure the MR for a field of $100\ \mbox{mT}$  applied approximately along the direction of the constriction ($70^\circ < \alpha < 110^\circ$) with the magnetization aligned along the angle $\alpha$. 
After removing the field, the spin orientation is determined by the shape anisotropy of the narrow structure causing the spins to align parallel to the edge. 
A head-to-head DW is formed at the constriction, associated with a resistance value of the nanocontact of $R_{AP} \sim 19\ \mbox{k}\Omega$. 
When the field is reversed, the magnetization configuration changes to a quasi-single domain state without a DW  ($R_P \sim 25\ \mbox{k}\Omega$) \cite{klaeui_domain_2003}. At higher reversed field, a new DW is nucleated.

Initially, all of the nominally identical samples exhibit a resistance of $\sim 275\ \Omega$ (Fig. \ref{fig2}(a)). For such low resistance values the MR is dominated by anisotropic magnetoresistance (AMR) with a magnitude of approximately 1\% \cite{patra_magnetoresistance_2010}. Up to $R_P \approx 1\ \mbox{k}\Omega$, the MR gradually increases due to the growing contribution of the constriction resistance to the total resistance. 
As discussed in our earlier study \cite{patra_magnetoresistance_2010}, the MR behavior of nanocontacts in this low resistance regime can be entirely explained by the bulk AMR effect. 
In contrast to what has been reported elsewhere \cite{bolotin_ballistic_2006}, we do not observe any sign of measurable DWMR in this low resistance regime and we can easily distinguish between DWMR and AMR from the angular dependences of the effect \cite{suppl_material}.

Next we turn to the ballistic conduction regime above $\sim 5\ \mbox{k}\Omega$, where novel MR effects are expected to occur. As the diameter of the magnetic nanocontact approaches atomic dimensions, thermal and electromigration effects can lead to significant rearrangements on the atomic scale, changing the total resistance of the contact. In contrast to the low resistance regime, the resistance changes during electromigration occur as distinct steps between stable levels as well-defined atomic reconfigurations take place at the narrowest part of the contact \cite{hoffmann_conductance_2008}. 
The MR changes significantly in this atomic contact regime. Its magnitude increases 
to more than $50\%$ (Fig. \ref{fig2}(a)) and we observe positive and negative MR. 
This large MR effect completely supersedes the small AMR \cite{ben_hamida_positive_2011} and dominates the overall MR.

Importantly, the switching fields in the R(H) loops do not change significantly during the thinning of the nanocontact, confirming that the magnetic states are fundamentally identical in both conduction regimes. As depinning fields usually depend strongly on the geometry and size of the constriction, a constant switching field indicates that the transition from the state with a DW to the one without does not occur by depinning of the wall. Instead, a reverse domain nucleates at one of the two ends of the half ring, which then annihilates the DW at the constriction. In contrast to previous studies (e.g. Refs. \cite{doudin_ballistic_2008,bolotin_ballistic_2006,sokolov_quantized_2007,egle_magnetoresistance_2010,ben_hamida_positive_2011, yamada_magnetoresistance_2011}), here the magnetization of the arms can be switched in low 
fields, independently of the precise geometry of the constriction, resulting in distinct stable configurations that can be probed at zero field. This means that we can uniquely identify the presence of a DW and the resulting impact on the MR even at zero field, as depicted in Fig. \ref{fig2}(c).

It was shown previously that magnetostriction artifacts can lead to arbitrarily high MR values in applied fields \cite{garcia_magnetoresistance_1999,chopra_ballistic_2002,egelhoff_artifacts_2004}. 
In contrast to most other studies, our nanocontacts are rigidly attached to the substrate and atomic force microscopy (AFM) as well as SEM imaging do not reveal significant suspended parts of the contacts. In addition, permalloy is known to exhibit extremely low magnetostriction leading to fm maximum length changes, which do not lead to significant resistance changes [see Fig. 3(d) in Ref. \onlinecite{achilles_tailoring_2011}]. We therefore do not expect significant effects from magnetostriction. 
To completely rule out magnetostriction due to externally applied fields and to gauge the applicability of the effect for non-volatile devices, we compare the MR values at \textit{zero} applied field: we obtain two distinct resistance levels with an MR ratio of up to 50\% (see for example Fig. \ref{fig2}(c)). Furthermore, apart from the switching event, the resistance levels do not change significantly when a field is applied (\ref{fig2}(b),(c)). For example, we do not observe any change in the resistance for fields above $70\ \mbox{mT}$ indicating that magnetostriction-related effects do not significantly contribute to the observed resistance change. 
At the same time, we do not observe a signature of tunneling transport (non-linear I-V characteristics or negative dR/dV \cite{bolotin_ballistic_2006}) in the interesting resistance range of 10-30 $ \mbox{k}\Omega$. This implies that the metallic contact is in the ballistic conduction regime and tunneling magnetoresistance (TMR) 
can be excluded as the dominating effect. In agreement with Ref. \onlinecite{ben_hamida_positive_2011}, our data shows that we can also exclude AMR since neither the angle-dependence nor the sign of the MR signal in our R($\alpha$) measurements agrees with the characteristics of 
 large AMR as sometimes observed in nanocontacts \cite{autes_giant_2008} (also see Fig. 1 in supplementary material \cite{suppl_material}). As we are in the ballistic conduction regime we can also exclude tunneling AMR as a source of the effects \cite{bolotin_anisotropic_2006}.

Ultimately, this 
indicates that it is the presence of a narrow DW in the ballistic transport regime that leads to this large MR. We can therefore conclude that we unambiguously observe DWMR.
Given the atomic size of the constriction in this resistance regime, it is the spin structure of the atoms at the very center of the contact, where the DW is located, that gives rise to the significant resistance changes observed. We have also studied pure Ni and Co contacts where we find similar results, indicating that the concurrent presence of Fe and Ni atoms in Py is not responsible for the observed effects.

 \begin{figure}[b]
 \includegraphics{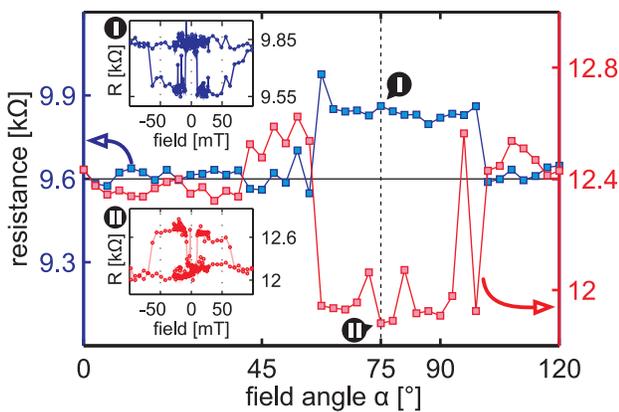}%
 \caption{\label{fig3} (Color online) Resistance of a nanocontact as a function of field angle $\alpha$ after applying a magnetic field and relaxing it to zero along $\alpha$. The data was obtained for two consecutive resistance states of a nanocontact (blue and red), indicating a change in the sign of MR associated with a change in contact resistance. Insets: R(H) major loops measured at $\alpha = 75^\circ$  obtained for the same contact configurations.}
 \end{figure}

Numerous models treat DWMR in the diffusive limit \cite{levy_resistivity_1997,tatara_resistivity_1997,van_gorkom_negative_1999,viret_spin_1996,wickles_electronic_2009,marrows_spin-polarised_2005} but few have considered the ballistic conduction regime. The reduced dimensions of such contacts require a self-consistent calculation of both the magnetic as well as the electronic structure of the nanocontact. 
Recent detailed ab initio calculations of this kind by Czerner et al. \cite{czerner_role_2010} yield DWMR values of around 50\% in line with our experimental observation. Jacob et al. \cite{jacob_magnetic_2005} conclude that realistic MR values in Ni nanocontacts are of the order of 30\%, similar to what we observe.

A key observation in our experiments is the occurrence of a sign change in the MR for a number of contacts in the ballistic regime (see Fig. \ref{fig2}(a)). This sign change is in contrast to previous experimental observations 
where such behavior was only found in the diffusive and in the tunneling regime \cite{bolotin_ballistic_2006}.  
Fig. \ref{fig3} shows the quasi-static MR for two consecutive resistance states of a nanocontact: While the resistance changes from $12\ \mbox{k}\Omega$ to $9.5\ \mbox{k}\Omega$, the MR jumps from $-4\%$ to $+3\%$. The corresponding R(H) loops (shown as insets) also show this behavior. 
Despite this change in resistance, likely caused by a small atomic reconfiguration at the narrowest part of the contact, the switching fields between the P and AP states remain at the same field values, allowing us to identify the magnetization configurations as explained above. Furthermore, the simultaneous occurrence of a resistance change and a sign change of the MR points to the same origin of the two effects. 

We therefore conclude that the underlying MR associated with the presence of a DW depends on the precise atomic configuration of the constriction. Theoretical approaches that are limited to simple geometries of the constriction, such as single-atomic wires, 
cannot satisfyingly describe this situation. 
Very recently Achilles et al. \cite{achilles_tailoring_2011} have considered different atomic configurations with resistance values similar to the ones observed in our experiments. Based on spin-dependent density functional theory, the MR is evaluated as the difference in resistance taken with and without a DW. Some of the atomic configurations considered differ only by the position of one or a few atoms. Precisely such atomic rearrangements can be induced by thermal activation, for instance during electromigration. In Ref. \onlinecite{achilles_tailoring_2011} the authors predict that, depending on the chosen configuration, the MR can be positive or negative with a strongly varying magnitude. This surprising result (in line with our observation) can be understood based on symmetry considerations: It is shown that a reduction of the symmetry of the nanocontact drastically reduces the conduction through the majority channel in the P state $g^{\uparrow\uparrow}_{P}$. In contrast to that, $g^{\downarrow\downarrow}_{P}$ and the conduction values in the AP state $g^{\uparrow\downarrow / \downarrow\uparrow}_{AP}$ are much less affected. Depending on the specific symmetries of the contact, this behavior is shown to cause $g_{AP} > g_{P}$ (negative MR) or $g_{AP} < g_{P}$ (positive MR). This result corroborates the hypothesis that small changes in the configuration of the nanocontact, observed as changes in the measured resistance, lead to pronounced changes in the MR including sign changes. Ultimately, this agreement between theory and experiment suggests that we can attribute the large MR changes to spin-dependent transport through discrete conductance channels that change their transmission depending on the atomic arrangement and the magnetic configuration of the narrowest part of the nanocontact. 

In summary, we have found a large DWMR (up to $50\%$) with both positive and negative sign at zero applied field in electromigrated Py nanocontacts in the ballistic transport regime. 
Our sample design and measurement scheme allow us to control the spin structure at the constriction. At the same time we are able to minimize artifacts due to field-induced magnetostriction and impurities, as our thorough analysis of possible artifacts shows. 
For the first time in the ballistic conduction regime, we demonstrate that the reproducible resistance states observed at zero applied magnetic field are associated with two stable magnetic configurations, with and without a DW pinned at the 
nanocontact. Our comparison of the measured MR behavior with available theoretical models shows that our results can only be reproduced by models that take into account spin polarized transport effects as well as the spatial and magnetic configuration of the atoms at the narrowest part of the nanocontact. Small changes in the atomic configuration, which appear as abrupt changes in the resistance, lead to a large change of the magnitude and even the sign of the MR that we observe. 
This means that both the sign and magnitude of the DWMR are governed by the precise geometrical arrangement of the constriction on the atomic scale, which so far has mostly been neglected.   

\begin{acknowledgments}
We acknowledge the financial support by the DFG (SFB 767, KL1811), the Swiss National Science Foundation, the ERC (ERC 2007-Stg 208162 and No. 2009-Stg 239838), the EU (RTN Spinswitch, MRTN CT-2006-035327), the Samsung Advanced Institute of Technology, and the Kompetenznetz Funktionelle Nanostrukturen Landesstiftung Baden W\"urttemberg. We thank S. Verleger for assistance with the deposition of the Au films, M. Mawass and A. Loescher for help with the analysis and discussions, and Prof. I. Mertig and S. Achilles for fruitful scientific discussions and sharing results of their theoretical studies.
\end{acknowledgments}

\bibliography{references_final}

\end{document}